\documentclass[aps,showpacs,preprintnumbers,amsmath,amssymb,twocolumn,superscriptaddress,prl]{revtex4-1}
\usepackage{txfonts}
\usepackage{float}
\usepackage{graphicx}
\usepackage{dcolumn}
\usepackage{bm}
\usepackage{hyperref}
\usepackage{subfigure}  

\def\comment#1{}

\def\slashchar#1{\setbox0=\hbox{$#1$}           
	\dimen0=\wd0                                 
	\setbox1=\hbox{/} \dimen1=\wd1               
	\ifdim\dimen0>\dimen1                        
	\rlap{\hbox to \dimen0{\hfil/\hfil}}      
	#1                                        
	\else                                        
	\rlap{\hbox to \dimen1{\hfil$#1$\hfil}}   
	/                                         
	\fi}                                         %

\begin{document}

\title{Analytical proof of Schottky Conjecture for multi-stage field emitters}

\author{Edgar Marcelino}
\email{edgarufba@gmail.com}
\address{Departamento de Física, Universidade Federal de Minas Gerais, C. P. 702, 30123-970, Belo Horizonte, MG, Brazil}

\begin{abstract}
Schottky Conjecture is analytically proved for multi-stage field emitters consisting on the superposition of rectangular or trapezoidal protrusions on a line under some specific limit. The case in which a triangular protrusion is present on the top of each emitter is also considered as an extension of the model. The results presented here are obtained via Schwarz-Christoffel conformal mapping and reinforce the validity of Schottky Conjecture when each protrusion is much larger than the ones above it, even when an arbitrary number of stages is considered. Moreover, it is showed that it is not necessary to require self-similarity between each of the stages in order to ensure the validity of the conjecture under the appropriate limits.
\end{abstract}

\pacs{85.45.Db, 85.45.Bz, 85.85.+j}
\maketitle

\section{Introduction}

One of the great achievements of Quantum Mechanics to Solid State Physics is to provide theoretical explanation for the emission of electrons from surfaces when a strong electrostatic field is applied \cite{Jeffreys,FowlerNordheim,Burgess,MurphyGood}, a process usually referred as Field Emission. This is an old topic of research but still full of important opened questions concerning the nature and modeling of this phenomenon \cite{ForbesFundmentals,JensenPRST,Holgate,Nanomaterials}. Besides the scientific and academic purposes, Field Emission is a topic of fundamental importance to technological developments and applications \cite{Mueller1,Mueller2,Mueller3,Ultramicroscopy,ColeBook}, specially in the cold field emission regime \cite{Jeffreys,FowlerNordheim,Burgess,MurphyGood,ForbesCFE1,ForbesCFE2}. Some of these applications include field emission displays, vacuum micro and nanoelectronics, satellite subsystems, mass spectrometers and even electrodynamics space tethers \cite{Colgan,Xu}. The theory of Cold Field Electron Emission was recently reformulated \cite{ForbesCFE1,ForbesCFE2,ForbesCFE3}, allowing for a better connection to experimental data. This includes the case of materials promising to emit electrons even under low applied electrostatic fields \cite{Han}.

In particular, Single Tip Field emitters (STFE) \cite{MarcelinoJVSTB,MarcelinoJAP,MarcelinoPRApplied,STFE1,STFE2,STFE3,STFE4} are of greater interest. This happens because strong electric fields are required to extract electrons from surfaces, usually from the order of a few volts per nanometer. Thus, geometries requiring corners, edges and tips are often studied, since they provide higher values of Field Enhancement Factor (FEF), defined in the following equation by the ratio between the absolute value of the electric field at some point $\mathbf{r}$ on the surface of the emitter, $|\mathbf{E(r)}|$, and the absolute value of the applied electric field, $|\mathbf{E_{0}}|$:
\begin{equation}
\gamma (\mathbf{r})=\frac{|\mathbf{E(r)}|}{|\mathbf{E_{0}}|}.
\end{equation}
 For this reason, applications involving STFE with very large aspect-ratios are often used. Nevertheless, STFEs with large aspect-ratios may involve some limitations due to the lack of mechanical stability. This problem may be solved by considering multi-stage structures, since these are able to provide mechanical stability together with significant increasing of the FEF, leading to intense theoretical investigation \cite{STFE1,STFE2,Schottky,Miller1,Miller2,MarcelinoJVSTB,AIPJensen}. Some times multi-stage emitters appear even at unexpected situations. For instance, the growth of nanotubes on carbon cloth may lead to extremely high values of FEF, such as 18800, obtained via the Fowler-Nordheim plots for Large Area Field Emitters (LAFE) \cite {WangAPL2004}. The mechanism leading to these giant values of FEF remained obscure until a deep study with Transmission Electron Microscope (TEM) reveal that the fibers grown on carbon cloth via thermal Chemical Vapor Deposition (CVD) may feature a multi-stage structure \cite{HuangTEM}.

In general, it is not trivial to determine the FEF of a multi-stage structure. Nevertheless, in 1923 a conjecture proposed by Schottky states that the FEF of a multi-stage structure would be the product of the FEFs of each of the individual stages \cite{Schottky}.  It is well known that this conjecture, often referred in the literature as Schottky Conjecture (SC), is not correct in general, although it predicts a very good approximation for the FEF of a multi-stage emitter when each of the protrusions is much smaller than the ones in which they are placed on. Indeed, SC is analytically proved to be valid under these limits in the case of a two-stage field emitter consisting of a rectangular protrusion placed on the center and the top of another rectangular protrusion on a line \cite{Miller1}. Latter, SC was also analytically proved for the case in which a triangular protrusion is placed on the center and the top of a rectangular one on a line, under the same limits \cite{MarcelinoJVSTB}, a result pointing out that the lack of self-similarity does not affect the validity of SC under these conditions. Surprisingly, different surveys based in many different methods have also verified the validity of SC at some significant region beyond the aforementioned situation \cite{STFE1,STFE2,Schottky,Miller1,Miller2,MarcelinoJVSTB,AIPJensen}. 

By using physical intuition, it might be expected that SC should be valid when a protrusion of very small dimensions acts like a perturbation placed over a much larger one. This happens because the smaller protrusion will experience, with very good approximation, the external field provided by the larger protrusion under the applied electrostatic field, which will not be significantly affected by the smaller protrusion, that simply represents a negligible perturbation. This logical path naturally yields SC as an expected result. Despite the simplicity of this argument, there is no general proof of SC, except for the very particular cases presented in \cite{Miller1,MarcelinoJVSTB}. Moreover, there is no analytical demonstration of SC for multi-stage field emitters even in this situation, although there are interesting evidences obtained from techniques using charge-models \cite{AIPJensen,Jensen_Verif,Jensen_Invest}. Indeed, there is no analytical proof of SC even in the case of the superposition of two cylindrical protrusions, the case originally studied by Schottky when he proposed his conjecture \cite{Schottky}. Nevertheless, some point- or line-charge models are able to provide shapes of emitters very similar to the case of ideal geometries, such as superimposed spherical, elliptical or cylindrical protrusions, and it is proved that  these models may provide the FEFs predicted by SC for the idealized structures under certain limits \cite{Jensen_Verif,Jensen_Invest}. 

Recent results obtained in Ref. \cite{Jensen_Invest} with line-charge models provide less than 70\% of agreement with SC even under the limits in which the conjecture is expected to be valid. It is possible that these results may be a consequence of the limitations of the techniques used, which provide a deformation of the stages when they are superimposed, compared to the single-stage situation, yielding this blatant disagreement with SC. Notwithstanding, the survey in Ref. \cite{Jensen_Invest} also points for the need of investigation and more general proofs of SC even under the limits in which the conjecture would be expected to be trivial. The present work intends to provide some contribution in this direction.

In the present manuscript SC is analytically proved via Schwarz-Christoffel conformal mapping \cite{SCT1,SCT2} for an arbitrary number of stages consisting of isosceles trapezoidal or rectangular protrusions, such that each protrusion is placed on the center of the top of another protrusion and has dimensions much smaller than the ones from the lower stage. For simplicity of the calculations, high aspect-ratios are also required.  Finally, the case in which a triangular protrusion is placed on the center of the top of these structures is also considered. By proving SC for an arbitrary number of stages with some generic (trapezoidal or triangular) shape, this work expects to reinforce the validity of SC for arbitrary shapes under the apropriate limits, since the (n-1) upper stages in a n-stage emitter can be viewed as a two-stage structure with some generic shape of the upper protrusion.

In section II  SC is proved for a multi-stage emitter compounded of rectangular protrusions, one placed on the center of the top of the lower one, such that the dimensions of any protrusion are much smaller than the dimensions from the lower one and each protrusion has high aspect-ratio. In Section III this demonstration is done for the case in which a triangular protrusion with high aspect-ratio and much smaller dimensions is placed on the center of the top of the upper protrusion. In Section IV the evaluation in section II is repeated for the case of isosceles trapezoidal stages and in Section V the case of Section IV is considered with a triangular protrusion of much smaller dimensions and high aspect-ratio centered in the apex. In Section VI a summary of the results and the conclusions are presented.


\section{Proof of SC for multiple rectangular stages}

\begin{figure}[H]
	\centering    
	\includegraphics[width=0.45\textwidth]{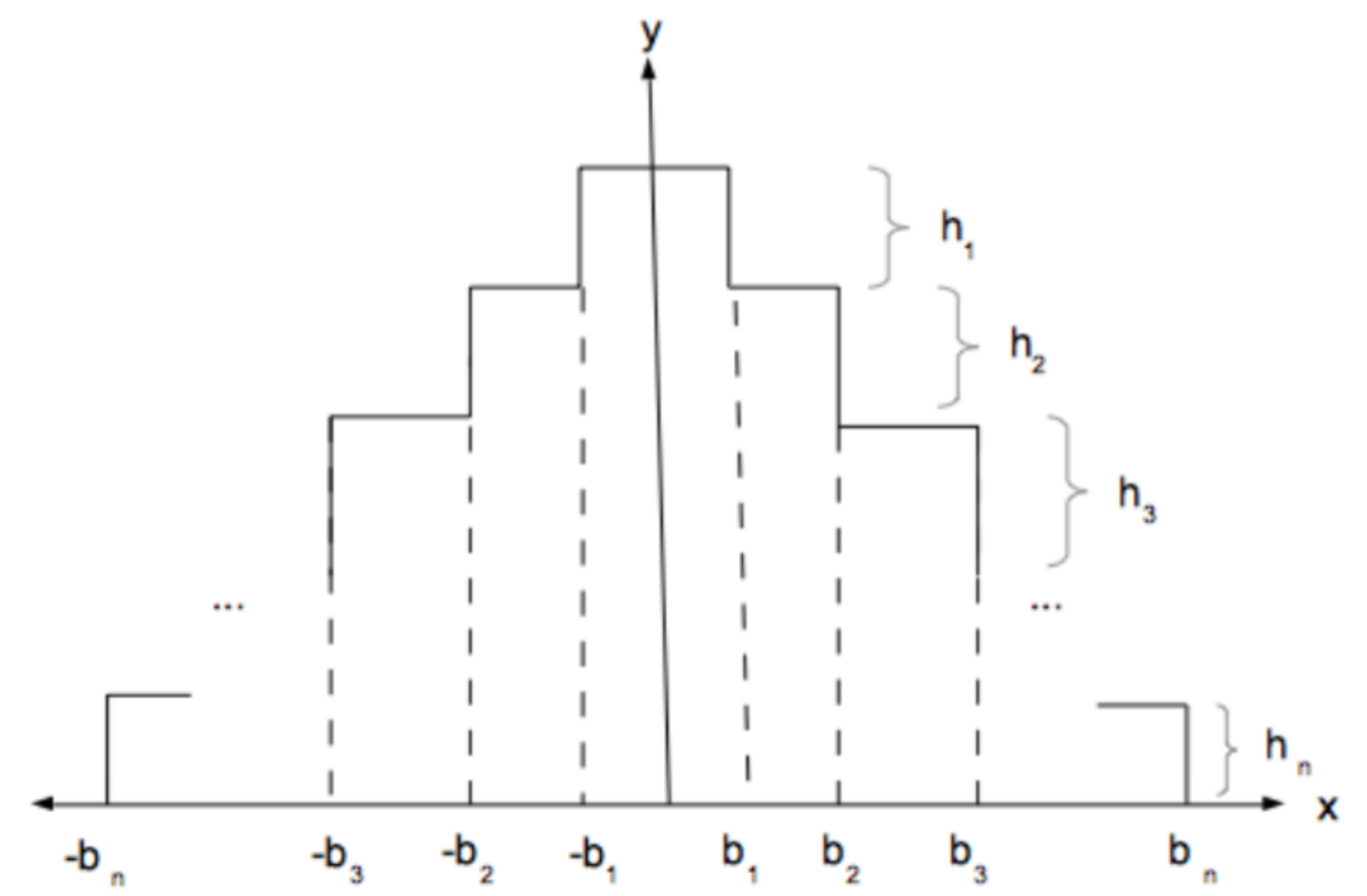}
	\caption{Emitter compounded of an arbitrary number, n, of rectangular protrusions of height $h_{j}$ and half-width $b_{j}$ ($j \in \{1,2,...,n\}$) .}	\label{Rect}
\end{figure}
Consider at first a two-dimensional emitter with the geometry determined by the polygonal line displayed on the z=x+iy$\rightarrow$(x,y)-plane in Fig. \ref{Rect}, consisting of n central rectangular protrusions of height $h_{j}$ and half-width $b_{j}$, with $j \in \{1,2,...,n\}$. The Schwarz-Christoffel transformation mapping the u-axis in the w=u+iv$\rightarrow$(u,v)-plane into the polygonal line in Fig. \ref{Rect} is given by the expression
\begin{equation}
z(w)=A \int_{u_{0}}^{w} \sqrt{\prod_{j=1}^{n}\left[\frac{w^{2}-u_{j}^{2}}{w^{2}-v_{j}^{2}} \right]}dw+B,
\end{equation}
where the parameters $u_j$ and $v_j$ satisfy the following correspondences between points in the w- and z-planes $(j \in \{1,2,...,n \})$: $z(w=u_{j})=b_{j}+i \sum_{k=j}^{n}h_{k}$ and $z(w=v_{j})=b_{j}+i \sum_{k=j+1}^{n}h_{k}$. Based on the Riemann Mapping Theorem \cite{Riemann}, one is free to choose three pre-images of the conformal mapping. Since the region outside the emitter can be viewed as a polygon with one vertex at infinity, there are two other arbitrary pre-images to choose \cite{MarcelinoErratum}. For instance, one can choose $z(w=u_{0} \equiv 0)=iH$, where $H \equiv \sum_{j=1}^{n} h_{j}$, and $u_{1}=1$. The correspondences for $w=u_{0} \equiv 0$ and $w=u_{1} \equiv 1$ lead to the equations
\begin{align}
B=iH, \\
A=\frac{b_{1}}{\int_{0}^{1} \sqrt{\prod_{k=1}^{n}\left[\frac{u_{k}^{2}-w^{2}}{v_{k}^{2}-w^{2}} \right]} dw}.
\end{align}
 Finally, these results together with the remaining correspondences lead to the following system of equations:
\begin{multline} \label{Sist1}
\frac{b_{j}-b_{j-1}}{b_1}=\frac{\int_{v_{j-1}}^{u_{j}} \sqrt{\prod_{k=1}^{j-1}\left[\frac{w^{2}-u_{k}^{2}}{w^{2}-v_{k}^{2}} \right] \prod_{l=j}^{n}\left[\frac{u_{l}^{2}-w^{2}}{v_{l}^{2}-w^{2}} \right]} dw}{ \int_{0}^{1} \sqrt{\prod_{k=1}^{n}\left[\frac{u_{k}^{2}-w^{2}}{v_{k}^{2}-w^{2}} \right]} dw} \quad (j \geq 2) , \\
\frac{h_{j}}{b_1}=\frac{\int_{u_{j}}^{v_{j}} \sqrt{ \left(\frac{w^{2}-u_{j}^{2}}{v_{j}^{2}-w^{2}} \right) \prod_{k=1}^{j-1}\left[\frac{w^{2}-u_{k}^{2}}{w^{2}-v_{k}^{2}} \right] \prod_{l=j+1}^{n}\left[\frac{u_{l}^{2}-w^{2}}{v_{l}^{2}-w^{2}} \right]} dw}{ \int_{0}^{1} \sqrt{\prod_{k=1}^{n}\left[\frac{u_{k}^{2}-w^{2}}{v_{k}^{2}-w^{2}} \right]} dw} \quad (j \geq 1),
\end{multline}
where the first equation comes from the correspondences for $u_{j}$ $(j \geq 2)$ and the second equation comes from the correspondences for $v_{j}$ $(j \geq 1)$.

The FEF on the top middle of the upper protrusion $(w=0)$ is given by 
\begin{equation}  \label{FEF1}
\gamma=\prod_{j=1}^{n} \left[ \frac{v_j}{u_j} \right].
\end{equation} 
Thus, the remaining task in order to determine this FEF is to solve the system of integral equations displayed in Eq. (\ref{Sist1}) for the parameters $u_j$ and $v_j$. This is a hard task that becomes analytically feasible in the limit $b_{1}<<h_{1}<<b_{2}<<h_{2}<<...<<b_{n}<<h_{n} \Rightarrow u_{1}<<v_{1}<<u_{2}<<v_{2}<<...<<u_{n}<<v_{n}$. Under this limit the following approximations are valid for the integrals in Eq. (\ref{Sist1}):

\begin{multline}
\int_{0}^{1} \sqrt{\prod_{k=1}^{n}\left[\frac{u_{k}^{2}-w^{2}}{v_{k}^{2}-w^{2}} \right]} dw \approx \frac{u_{2}u_{3}...u_{n}}{v_{1}v_{2}...v_{n}} \int_{0}^{1} \sqrt{1-w^{2}} dw= \\
=\frac{\pi}{4} \frac{u_{2}u_{3}...u_{n}}{v_{1}v_{2}...v_{n}} ,
\end{multline}
\begin{multline}
\int_{v_{j-1}}^{u_{j}} \sqrt{\prod_{k=1}^{j-1}\left[\frac{w^{2}-u_{k}^{2}}{w^{2}-v_{k}^{2}} \right] \prod_{l=j}^{n}\left[\frac{u_{l}^{2}-w^{2}}{v_{l}^{2}-w^{2}} \right]} dw \approx \\
\approx \frac{u_{j+1}u_{j+2}...u_{n}}{v_{j}v_{j+1}...v_{n}} \int_{v_{j-1}}^{u_{j}} w \sqrt{\frac{u_{j}^{2}-w^{2}}{w^{2}-v_{j-1}^{2}}} dw = \\
 =\frac{u_{j+1}u_{j+2}...u_{n}}{v_{j}v_{j+1}...v_{n}} \int_{v_{j-1}}^{u_{j}} w \sqrt{\frac{1-(w^{2}/u_{j}^{2})}{(w^{2}/u_{j}^{2})-(v_{j-1}^{2}/u_{j}^{2})}} dw \approx \\
 \approx  \frac{u_{j+1}u_{j+2}...u_{n}}{v_{j}v_{j+1}...v_{n}} \int_{v_{j-1}}^{u_{j}} \sqrt{u_{j}^{2}-w^{2}} dw= \\
 = \frac{u_{j+1}u_{j+2}...u_{n}}{v_{j}v_{j+1}...v_{n}} \left[\frac{u_{j}^{2}}{2} \arcsin \left( \frac{w}{u_j} \right)+\frac{w\sqrt{u_{j}^{2}-w^{2}}}{2} \right]_{v_{j-1}}^{u_j} \approx \\
 \approx \frac{u_{j+1}u_{j+2}...u_{n}}{v_{j}v_{j+1}...v_{n}} \frac{\pi u_j^{2}}{4},
\end{multline}
\begin{multline}
\int_{u_{j}}^{v_{j}} \sqrt{ \left(\frac{w^{2}-u_{j}^{2}}{v_{j}^{2}-w^{2}} \right) \prod_{k=1}^{j-1}\left[\frac{w^{2}-u_{k}^{2}}{w^{2}-v_{k}^{2}} \right] \prod_{l=j+1}^{n}\left[\frac{u_{l}^{2}-w^{2}}{v_{l}^{2}-w^{2}} \right]} dw \approx \\
\approx \frac{u_{j+1}u_{j+2}...u_{n}}{v_{j+1}v_{j+2}...v_{n}} \int_{u_{j}}^{v_{j}} \sqrt{\frac{w^{2}-u_{j}^{2}}{v_{j}^{2}-w^{2}}}dw= \\
\frac{u_{j+1}u_{j+2}...u_{n}}{v_{j+1}v_{j+2}...v_{n}} \int_{u_{j}}^{v_{j}} \sqrt{\frac{v_{j}^{2}-u_{j}^{2}}{v_{j}^{2}-w^{2}}-1}dw \approx \\
\approx \frac{u_{j+1}u_{j+2}...u_{n}}{v_{j+1}v_{j+2}...v_{n}} \int_{u_{j}}^{v_{j}} \sqrt{\frac{v_{j}^{2}}{v_{j}^{2}-w^{2}}-1}dw= \\
=\frac{u_{j+1}u_{j+2}...u_{n}}{v_{j+1}v_{j+2}...v_{n}} \int_{u_{j}}^{v_{j}} \frac{w dw}{\sqrt{v_{j}^{2}-w^{2}}}= \\
=\frac{u_{j+1}u_{j+2}...u_{n}}{v_{j+1}v_{j+2}...v_{n}} \sqrt{v_{j}^{2}-u_{j}^{2}}
\approx \frac{u_{j+1}u_{j+2}...u_{n}}{v_{j+1}v_{j+2}...v_{n}} v_{j}.
\end{multline}

With all these approximations, Eq. (\ref{Sist1}) reduces to the following system;
\begin{align} \label{Ratio1}
\frac{b_{j}}{b_1}=\frac{v_{1}v_{2}...v_{j-1}}{u_{2}u_{3}...u_{j-1}}u_{j} \\
\frac{h_{j}}{b_1}=\frac{4}{\pi} \frac{v_{1}v_{2}...v_{j}}{u_{1}u_{2}...u_{j}}v_{j},
\end{align}
that leads to the final equation
\begin{equation}
\frac{v_j}{u_j}=\sqrt{\frac{\pi h_{j}}{4b_{j}}}.
\end{equation}
Thus, after using this result in Eq. (\ref{FEF1}), the expression for the FEF on the top middle of the structure can be easily obtained:
\begin{equation} \label{Final1}
\gamma(x=0,y=H)=\prod_{j=1}^{n} \left[\sqrt{\frac{\pi h_{j}}{4b_{j}}} \right].
\end{equation}

Since the FEF on the center of the top of a single rectangular protrusion of height $h_1$ and half-width $b_1$ on a line is given by $\gamma_{1}=\sqrt{\frac{\pi h_{1}}{4b_{1}}}$, when the aspect-ratio of the protrusion is very large \cite{Miller1}, Eq. (\ref{Final1}) proves SC in the limit $b_1<<h_1<<b_2<<h_2<<...<<b_n<<h_n$ for an arbitrary number, n, of superimposed rectangular protrusions on a line. 

\section{Proof of SC for multiple rectangular stages with a triangular stage on the top}

\begin{figure}[H]
	\centering    
	\includegraphics[width=0.45\textwidth]{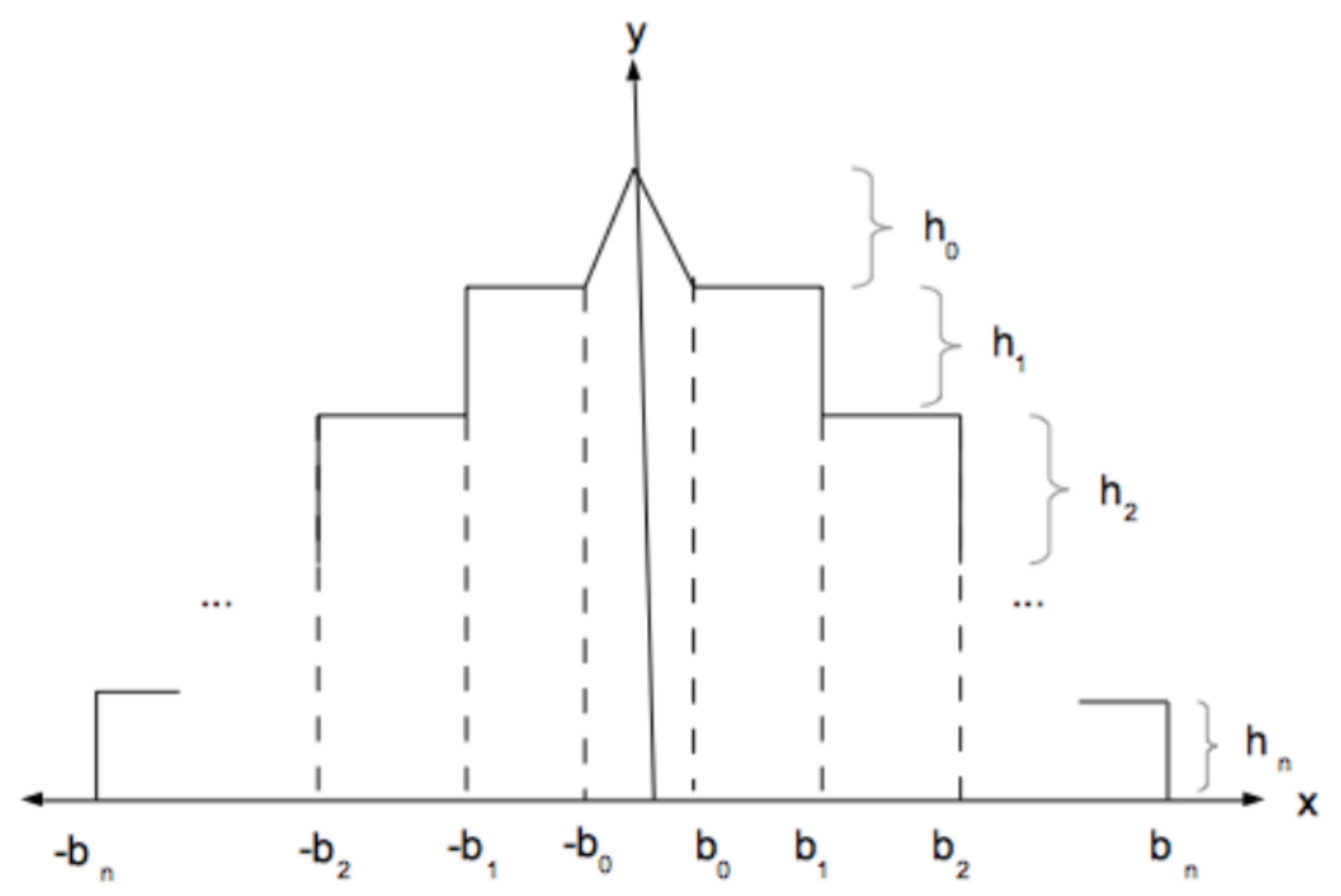}
	\caption{Emitter compounded of an arbitrary number, n, of rectangular protrusions of height $h_{j}$ and half-width $b_{j}$ ($j \in \{1,2,...,n\}$) with a triangular protrusion of height $h_{0}$ and half-width $b_{0}$ placed on the center of the top of the highest rectangular protrusion.}	\label{RectTriang}
\end{figure}

At this section it is considered the case in which there is an isosceles triangular protrusion on the center of the top of the highest rectangular protrusion, for the field emitter studied in the previous section, see Fig. \ref{RectTriang}. In this case, the Schwarz-Christoffel transformation mapping the u-axis in the $w=u+iv \rightarrow (u,v)$-plane into the polygonal line displayed in Fig. \ref{RectTriang} in the $z=x+iy \rightarrow (x,y)$-plane is given by:
\begin{equation} \label{SCT2}
z(w)=A \int_{u_0}^{w} \frac{w^{1-\alpha}dw}{(w^{2}-1)^{\frac{1-\alpha}{2}}} \sqrt{\prod_{j=1}^{n}\left[\frac{w^{2}-u_{j}^{2}}{w^{2}-v_{j}^{2}} \right]}+B,
\end{equation}
where $\alpha=\frac{2}{\pi} \arctan \left( \frac{b_0}{h_0} \right)$ corresponds to the internal angle from the upper vertex of the triangular protrusion. The identities $z(w=u_{0}=0)=iH$ and $z(w=\pm v_{0}=\pm 1)=\pm b_{0}+i(H-h_{0})$, are chose to be satisfied based on the Riemann Mapping Theorem \cite{Riemann} and $H$ is defined as $H=\sum_{j=0}^{n} h_{j}$. These identities lead to the expressions:
\begin{eqnarray}
B=iH, \\
A=\frac{\sqrt{b_{0}^{2}+h_{0}^{2}}}{\int_{0}^{1} \frac{w^{1-\alpha}dw}{(1-w^{2})^{\frac{1-\alpha}{2}}} \sqrt{\prod_{j=1}^{n}\left[\frac{u_{j}^{2}-w^{2}}{v_{j}^{2}-w^{2}} \right]}}. \label{ARectTriang}
\end{eqnarray}
For $j \in \{1,2,...,n\}$, the other correspondences from the Schwarz-Christoffel transformation, $z(w=u_j)=b_{j}+i (H-\sum_{k=0}^{j-1} h_{k})$ and $z(w=v_{j})=b_{j}+i(H-\sum_{k=0}^{j} h_{k})$, can be used together with the previous equations to obtain;
\begin{multline} \label{Sist2}
\frac{b_{j}-b_{j-1}}{\sqrt{b_{0}^{2}+h_{0}^{2}}}= \frac{\int_{v_{j-1}}^{u_{j}} \frac{w^{1-\alpha}dw}{(w^{2}-1)^{\frac{1-\alpha}{2}}}  \sqrt{\prod_{k=1}^{j-1}\left[\frac{w^{2}-u_{k}^{2}}{w^{2}-v_{k}^{2}} \right] \prod_{l=j}^{n}\left[\frac{u_{l}^{2}-w^{2}}{v_{l}^{2}-w^{2}} \right]}}{ \int_{0}^{1} \frac{w^{1-\alpha}dw}{(1-w^{2})^{\frac{1-\alpha}{2}}} \sqrt{\prod_{k=1}^{n}\left[\frac{u_{k}^{2}-w^{2}}{v_{k}^{2}-w^{2}} \right]}} , \\
\frac{h_{j}}{\sqrt{b_{0}^{2}+h_{0}^{2}}}=\frac{\int_{u_{j}}^{v_{j}} \frac{w^{1-\alpha}dw}{(w^{2}-1)^{\frac{1-\alpha}{2}}}   \sqrt{ \left(\frac{w^{2}-u_{j}^{2}}{v_{j}^{2}-w^{2}} \right) \prod_{k=1}^{j-1}\left[\frac{w^{2}-u_{k}^{2}}{w^{2}-v_{k}^{2}} \right] \prod_{l=j+1}^{n}\left[\frac{u_{l}^{2}-w^{2}}{v_{l}^{2}-w^{2}} \right]}}{ \int_{0}^{1} \frac{w^{1-\alpha}dw}{(1-w^{2})^{\frac{1-\alpha}{2}}}  \sqrt{\prod_{k=1}^{n}\left[\frac{u_{k}^{2}-w^{2}}{v_{k}^{2}-w^{2}} \right]}}.
\end{multline}
By solving the system in Eq. (\ref{Sist2}), one can find the parameters $u_{j}$ and $v_{j}$ ($j \in \{1,2,...,n\}$) and then obtain the FEF close to the triangular apex $(w \approx 0)$, given by:
\begin{equation}
\gamma (w)= \left( \prod_{j=1}^{n} \frac{v_j}{u_j} \right) \frac{1}{|w|^{1-\alpha}}.
\end{equation}
But near $w \approx 0$, Eq. (\ref{SCT2}) reduces to
\begin{equation}
|z-iH|=\left( \prod_{j=1}^{n} \frac{u_j}{v_j} \right) \frac{|w|^{2-\alpha}}{2-\alpha}.
\end{equation}
Thus, the final expression for the FEF will be
\begin{equation} \label{FEF2}
\gamma (w \approx 0)=\left( \prod_{j=1}^{n} \frac{v_j}{u_j} \right) \left[ \frac{A \left( \prod_{j=1}^{n} \frac{u_j}{v_j} \right)}{(2-\alpha) |z-iH|}\right]^{\frac{1-\alpha}{2-\alpha}}.
\end{equation}
Since $A$ is determined from Eq. (\ref{ARectTriang}), it remains to solve the system in Eq. (\ref{Sist2}) in order to obtain the parameters $u_j$ and $v_j$ in Eq. (\ref{FEF2}) and find the FEF expression for an arbitrary point $(x,y)$ close to the apex. Again an analytical solution of the system of equations is only feasible under some specific limit. This is the limit $b_{0}<<h_{0}<<b_{1}<<h_{1}<...<<b_{n}<<h_{n} \Rightarrow 1<<u_{1}<<v_{1}<<u_{2}<<_{v2}<<u_{n}<<v_{n}$. This assumption leads to the following approximtions, in analogy to the ones in the previous section:
\begin{multline}
\int_{0}^{1} \frac{w^{1-\alpha}dw}{(1-w^{2})^{\frac{1-\alpha}{2}}} \sqrt{\prod_{k=1}^{n}\left[\frac{u_{k}^{2}-w^{2}}{v_{k}^{2}-w^{2}} \right]} \approx \\
\approx \left(\prod^{n}_{j=1}\frac{u_{j}}{v_{j}} \right) \frac{\Gamma \left(1-\frac{\alpha}{2} \right) \Gamma \left(\frac{1+\alpha}{2} \right)}{\sqrt{\pi}} ,  \\
\int_{v_{j-1}}^{u_{j}} \frac{w^{1-\alpha}dw}{(w^{2}-1)^{\frac{1-\alpha}{2}}}  \sqrt{\prod_{k=1}^{j-1}\left[\frac{w^{2}-u_{k}^{2}}{w^{2}-v_{k}^{2}} \right] \prod_{l=j}^{n}\left[\frac{u_{l}^{2}-w^{2}}{v_{l}^{2}-w^{2}} \right]} \approx \\
\approx \left(\prod^{n}_{k=j+1} \frac{u_{k}}{v_{k}} \right) \frac{\pi u_{j}^{2}}{4} , \\
\int_{u_{j}}^{v_{j}} \frac{w^{1-\alpha}dw}{(w^{2}-1)^{\frac{1-\alpha}{2}}}   \sqrt{ \left(\frac{w^{2}-u_{j}^{2}}{v_{j}^{2}-w^{2}} \right) \prod_{k=1}^{j-1}\left[\frac{w^{2}-u_{k}^{2}}{w^{2}-v_{k}^{2}} \right] \prod_{l=j+1}^{n}\left[\frac{u_{l}^{2}-w^{2}}{v_{l}^{2}-w^{2}} \right]} \approx \\
\approx \left(\prod^{n}_{k=j+1} \frac{u_{k}}{v_{k}} \right) v_{j}
\end{multline}
Finally, Eq. (\ref{Sist2}) reduces to,
\begin{align}
\frac{b_j}{h_{0}}= \frac{u_{j}^{2}}{v_j} \left(\prod^{j}_{k=1} \frac{v_{k}}{u_{k}} \right) \frac{\pi^{3/2} }{4 \Gamma \left(1-\frac{\alpha}{2} \right) \Gamma \left(\frac{1+\alpha}{2} \right)} , \\
\frac{h_j}{h_{0}}= v_{j} \left(\prod^{j}_{k=1} \frac{v_{k}}{u_{k}} \right) \frac{\sqrt{\pi}}{ \Gamma \left(1-\frac{\alpha}{2} \right) \Gamma \left(\frac{1+\alpha}{2} \right)}.
\end{align}
Thus, a simple manipulation of the previous equations leads to;
\begin{equation}
\frac{v_j}{u_j}=\sqrt{\frac{\pi h_{j}}{4b_{j}}}.
\end{equation}
Besides that, Eq. (\ref{ARectTriang}) can be approximated to
\begin{equation}
A \left(\prod^{n}_{j=1} \frac{u_{j}}{v_{j}} \right) \approx \frac{h_{0} \sqrt{\pi}}{ \Gamma \left(1-\frac{\alpha}{2} \right) \Gamma \left(\frac{1+\alpha}{2} \right)}.
\end{equation}
Substituting these results in Eq. (\ref{FEF2}), the final expression for the FEF close to the apex of the emitter van be obtained:
\begin{multline} \label{Final2}
\gamma(x \approx 0,y \approx H)= \\
=\prod_{j=1}^{n} \left[\sqrt{\frac{\pi h_{j}}{4b_{j}}} \right] \left[ \frac{\sqrt{\pi}}{(2-\alpha) \frac{\xi(x,y)}{h_0}  \Gamma \left(1-\frac{\alpha}{2} \right) \Gamma \left(\frac{1+\alpha}{2} \right)}\right]^{\frac{1-\alpha}{2-\alpha}},
\end{multline}
where $\xi(x,y)=\sqrt{x^{2}+(y-H)^{2}}$.

Eq. (\ref{Final2}) shows that under the limit $b_{0}<<h_{0}<<b_{1}<<h_{1}<...<<b_{n}<<h_{n} \Rightarrow 1<<u_{1}<<v_{1}<<u_{2}<<_{v2}<<u_{n}<<v_{n}$, the FEF can be written as the product of the FEFs corresponding to each rectangular protrusion on a line, see Ref. \onlinecite{Miller1}, wth the FEF close to the apex of a single triangular protrusion on a line, see Ref. \onlinecite{MarcelinoJVSTB}. Thus SC is proved for this system under this limit.

\section{Proof of SC for multiple trapezoidal stages}

\begin{figure}[H]
	\centering    
	\includegraphics[width=0.45\textwidth]{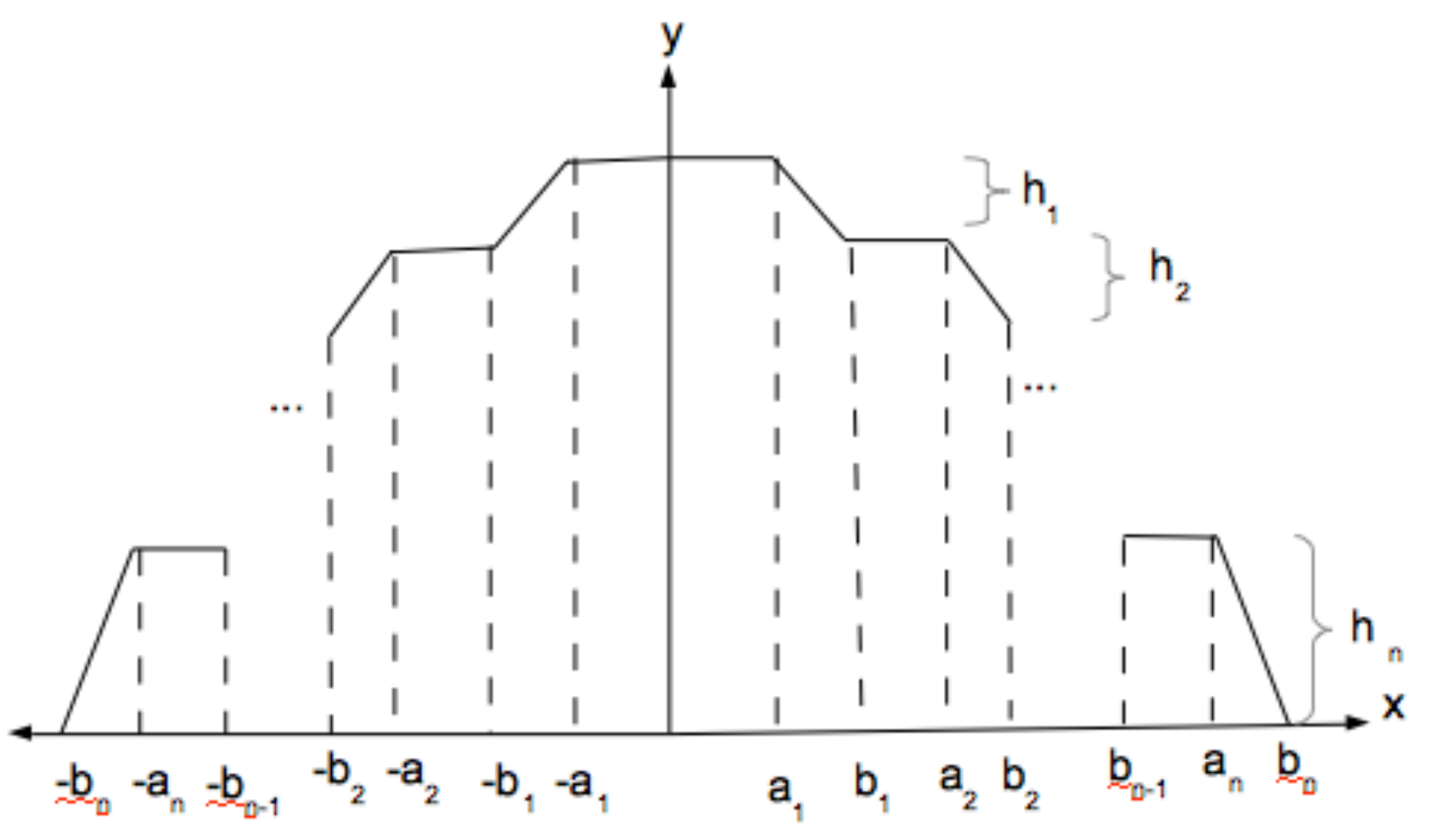}
	\caption{Emitter compounded of an arbitrary number, n, of trapezoidal protrusions of height $h_{j}$ and half-width from the lower base $b_{j}$ ($j \in \{1,2,...,n\}$). }	\label{Trap}
\end{figure}

In this section the FEF on the center of the top of an emitter consisting of $n$ superimposed trapezoidal stages is considered, see Fig. \ref{Trap}. The emitter is such that each trapezoidal protrusion is isoceles and placed on the center of the top of the other protrusion and each stage has dimensions much larger than the ones from the protrusions placed above. For simplicity of the calculation, only high aspect-ratios are considered. The Schwarz-Christoffel transformation mapping the u-axis in the w=u+iv$\rightarrow$(u,v)-plane into the polygonal line in Fig. \ref{Trap} is given by the expression
\begin{equation} \label{SCT3}
z(w)=A \int_{u_0}^{w} \prod_{j=1}^{n}\left[\frac{w^{2}-u_{j}^{2}}{w^{2}-v_{j}^{2}} \right]^{\alpha_{j}}dw+B,
\end{equation}
where $(j \in \{1,2,...,n\})$ $\alpha_{j}=(1/ \pi) \arctan \left( \frac{h_j}{b_{j}-a_{j}} \right)$ corresponds to the angles of the lower base of each trapeze with the adjacent sides and the parameters $u_j$ and $v_j$ satisfy the following correspondences $(j \geq 1)$ between points in the w- and z-planes: $z(w=u_{j})=a_{j}+i \sum_{k=j}^{n} h_{k}$ and $z(w=v_{j})=b_{j}+i \sum_{k=j+1}^{n}h_{k}$. Based on the Riemann Mapping Theorem \cite{Riemann}, one can choose $z(w=u_{0} \equiv 0)=iH$, where $H \equiv \sum_{j=1}^{n} h_{j}$, and $u_{1}=1$. The correspondences for $w=u_{0} \equiv 0$ and $w=u_{1} \equiv 1$ lead to the equations
\begin{align}
B=iH, \\
A =\frac{a}{\int_{0}^{1} \prod_{j=1}^{n}\left[\frac{u_{j}^{2}-w^{2}}{v_{j}^{2}-w^{2}} \right]^{\alpha_{j}}dw}.
\end{align}
 Finally, these results together with the remaining correspondences lead to the following system of equations:
\begin{multline} \label{Sist3}
\frac{a_{j}-b_{j-1}}{a_1}=\frac{\int_{v_{j-1}}^{u_{j}} \prod_{k=1}^{j-1}\left[\frac{w^{2}-u_{k}^{2}}{w^{2}-v_{k}^{2}} \right]^{\alpha_{k}} \prod_{l=j}^{n}\left[\frac{u_{l}^{2}-w^{2}}{v_{l}^{2}-w^{2}} \right]^{\alpha_{l}} dw}{ \int_{0}^{1} \prod_{k=1}^{n}\left[\frac{u_{k}^{2}-w^{2}}{v_{k}^{2}-w^{2}} \right]^{\alpha_{k}} dw}  , \\
\frac{1}{a_1}=\frac{\int_{u_{j}}^{v_{j}}  \left(\frac{w^{2}-u_{j}^{2}}{v_{j}^{2}-w^{2}} \right)^{\alpha_{j}}  \prod_{k=1}^{j-1}\left[\frac{w^{2}-u_{k}^{2}}{w^{2}-v_{k}^{2}} \right]^{\alpha_{k}}  \prod_{l=j+1}^{n}\left[\frac{u_{l}^{2}-w^{2}}{v_{l}^{2}-w^{2}} \right]^{\alpha_{l}} dw}{\sqrt{(b_{j}-a_{j})^{2}+h_{j}^{2}} \int_{0}^{1} \prod_{k=1}^{n}\left[\frac{u_{k}^{2}-w^{2}}{v_{k}^{2}-w^{2}} \right]^{\alpha_{k}} dw} ,
\end{multline}
where the first equation comes from the correspondences for $u_{j}$ $(j \geq 2)$ and the second equation comes from the correspondences for $v_{j}$ $(j \geq 1)$. These equations can be combined and lead to the following one:
\begin{multline} \label{equationuv}
\frac{a_{j}-b_{j-1}}{\sqrt{(b_{j}-a_{j})^{2}+h_{j}^{2}}}=  \\
=\frac{\int_{v_{j-1}}^{u_{j}} \prod_{k=1}^{j-1}\left[\frac{w^{2}-u_{k}^{2}}{w^{2}-v_{k}^{2}} \right]^{\alpha_{k}} \prod_{l=j}^{n}\left[\frac{u_{l}^{2}-w^{2}}{v_{l}^{2}-w^{2}} \right]^{\alpha_{l}} dw}{\int_{u_{j}}^{v_{j}}  \left(\frac{w^{2}-u_{j}^{2}}{v_{j}^{2}-w^{2}} \right)^{\alpha_{j}}  \prod_{k=1}^{j-1}\left[\frac{w^{2}-u_{k}^{2}}{w^{2}-v_{k}^{2}} \right]^{\alpha_{k}}  \prod_{l=j+1}^{n}\left[\frac{u_{l}^{2}-w^{2}}{v_{l}^{2}-w^{2}} \right]^{\alpha_{l}} dw}.
\end{multline}
In the limit $a_{1} \leq b_{1}<<h_{1}<<a_{2}\leq b_{2}<<h_{2}<<...<<a_{n}\leq b_{n}<<h_{n} \Rightarrow u_{1}<<v_{1}<<u_{2}<<v_{2}<<....<<u_{n}<<v_{n}$, the following approximations are valid:
\begin{multline}
\int_{v_{j-1}}^{u_{j}} \prod_{k=1}^{j-1} \left[\frac{w^{2}-u_{k}^{2}}{w^{2}-v_{k}^{2}} \right]^{\alpha_{k}} \prod_{l=j}^{n}\left[\frac{u_{l}^{2}-w^{2}}{v_{l}^{2}-w^{2}} \right]^{\alpha_{l}} dw  \approx \\
\approx \prod_{k=j+1}^{n} \left(\frac{u_k}{v_k}  \right)^{2 \alpha_{k}} \frac{1}{v_{j}^{2 \alpha_{j}}}   \int_{v_{j-1}}^{u_{j}} \frac{w^{2 \alpha_{j-1}} (u_{j}^{2}-w^{2})^{\alpha_{j}} dw}{(w^{2}-v_{j-1}^{2})^{\alpha_{j-1}}} = \\
= \prod_{k=j+1}^{n} \left(\frac{u_k}{v_k}  \right)^{2 \alpha_{k}} \frac{u_{j}^{2 \alpha_{j}+1}}{v_{j}^{2 \alpha_{j}}}   \int_{\frac{v_{j-1}}{u_{j}}}^{1} \frac{x^{2 \alpha_{j-1}} (1-x^{2})^{\alpha_{j}} dx}{(x^{2}-\frac{v_{j-1}^{2}}{u_{j}^{2}})^{\alpha_{j-1}}} \approx \\
\approx \prod_{k=j+1}^{n} \left(\frac{u_k}{v_k}  \right)^{2 \alpha_{k}} \frac{u_{j}^{2 \alpha_{j}+1}}{v_{j}^{2 \alpha_{j}}}   \int_{0}^{1}  (1-x^{2})^{\alpha_{j}} dx= \\
= \prod_{k=j+1}^{n} \left(\frac{u_k}{v_k}  \right)^{2 \alpha_{k}} \frac{u_{j}^{2 \alpha_{j}+1}}{2 v_{j}^{2 \alpha_{j}}}   \int_{0}^{1}  (1-x)^{\alpha_{j}} x^{-1/2} dx = \\
= \prod_{k=j+1}^{n} \left(\frac{u_k}{v_k}  \right)^{2 \alpha_{k}} \frac{u_{j}^{2 \alpha_{j}+1}}{2 v_{j}^{2 \alpha_{j}}} \frac{\sqrt{\pi} \Gamma(\alpha_{j}+1)}{\Gamma \left(\alpha_{j}+\frac{3}{2} \right)} ,
\end{multline} 
\begin{multline}
\int_{u_{j}}^{v_{j}}  \left(\frac{w^{2}-u_{j}^{2}}{v_{j}^{2}-w^{2}} \right)^{\alpha_{j}}  \prod_{k=1}^{j-1}\left[\frac{w^{2}-u_{k}^{2}}{w^{2}-v_{k}^{2}} \right]^{\alpha_{k}}  \prod_{l=j+1}^{n}\left[\frac{u_{l}^{2}-w^{2}}{v_{l}^{2}-w^{2}} \right]^{\alpha_{l}} dw \approx \\
\approx \prod_{k=j+1}^{n} \left(\frac{u_k}{v_k}  \right)^{2 \alpha_{k}} \int_{u_{j}}^{v_{j}}  \left(\frac{w^{2}-u_{j}^{2}}{v_{j}^{2}-w^{2}} \right)^{\alpha_{j}} dw= \\
=v_{j} \prod_{k=j+1}^{n} \left(\frac{u_k}{v_k}  \right)^{2 \alpha_{k}} \int_{\frac{u_{j}}{v_{j}}}^{1}  \left(\frac{w^{2}-\frac{u_{j}^{2}}{v_{j}^{2}}}{1-w^{2}} \right)^{\alpha_{j}} dw \approx \\
\approx  v_{j} \prod_{k=j+1}^{n} \left(\frac{u_k}{v_k}  \right)^{2 \alpha_{k}} \int_{0}^{1} \frac{w^{2 \alpha_{j}}dw}{(1-w^{2})^{\alpha_{j}}}= \\
= \frac{v_{j}}{2} \prod_{k=j+1}^{n} \left(\frac{u_k}{v_k}  \right)^{2 \alpha_{k}} \int_{0}^{1} x^{\alpha_{j}-\frac{1}{2}} (1-x)^{-\alpha_{j}}dx= \\
=\frac{v_{j} \Gamma(1-\alpha_{j}) \Gamma \left(\alpha_{j}+\frac{1}{2} \right)}{\sqrt{\pi}} \prod_{k=j+1}^{n} \left(\frac{u_k}{v_k}  \right)^{2 \alpha_{k}}.
\end{multline} 
Finally, Eq. (\ref{equationuv}) with these approximations leads to;
\begin{multline}
\frac{a_{j}-b_{j-1}}{\sqrt{(b_{j}-a_{j})^{2}+h_{j}^{2}}}=\frac{a_{j}-b_{j-1}}{h_{j}} \sin (\pi \alpha_{j}) \approx \frac{a_{j}}{h_{j}} \sin (\pi \alpha_{j}) \approx \\
\approx  \frac{ \pi \alpha_{j} \Gamma(\alpha_{j})}{2 \left( \alpha_{j}+\frac{1}{2} \right) \left[\Gamma \left(\alpha_{j}+\frac{1}{2} \right) \right]^{2} \Gamma(1-\alpha_{j})} \left(\frac{u_j}{v_j} \right)^{2 \alpha_{j}+1},
\end{multline}
which after the use of the Reflection Formula of the gamma function results:
\begin{equation} \label{Equv34}
\frac{v_{j}}{u_{j}} \approx \left[ \frac{ \alpha_{j} \left[\Gamma (\alpha_{j}) \right]^{2}}{(2 \alpha_{j}+1) \left[\Gamma \left(\alpha_{j}+\frac{1}{2} \right) \right]^{2} } \frac{h_j}{a_j}\right]^{1/(2 \alpha_{j}+1)}.
\end{equation}

From Eq. (\ref{SCT3}), it is easy to see that the FEF on the top middle of the structure $(w=0)$ can be expressed by
\begin{equation}
\gamma(x=0,y=H)=\prod_{j=1}^{n} \left(\frac{v_j}{u_j}  \right)^{2 \alpha_{j}} .
\end{equation}
Thus, the FEF on the center of the apex of the emitter is given by:
\begin{equation} \label{Final3}
\gamma(x \approx 0,y \approx H)=\prod_{j=1}^{n} \left[ \frac{ \alpha_{j} \left[\Gamma (\alpha_{j}) \right]^{2}}{(2 \alpha_{j}+1) \left[\Gamma \left(\alpha_{j}+\frac{1}{2} \right) \right]^{2} } \frac{h_j}{a_j}\right]^{2 \alpha_{j} /(2 \alpha_{j}+1)}
\end{equation}
Comparison with the FEF on the top middle of an isosceles trapezoidal protrusion with high aspect-ratio on a line \cite{Miller2} shows that the FEF evaluated in Eq. (\ref{Final3}) is given by the product of the FEFs of each of the n trapezoidal protrusions on a line. Thus, SC is proved for the emitter considered along this section under the previously mentioned approximations. As expected, Eq. (\ref{Final1}) is recovered from Eq. (\ref{Final3}) when $\alpha_{j}=1/2$ for all $j \in \{1,2,...,n\}$, since in this case the trapezoidal protrusions become rectangular.

\section{Proof of SC for multiple trapezoidal stages with a triangular stage on the top}

\begin{figure}[H]
	\centering    
	\includegraphics[width=0.45\textwidth]{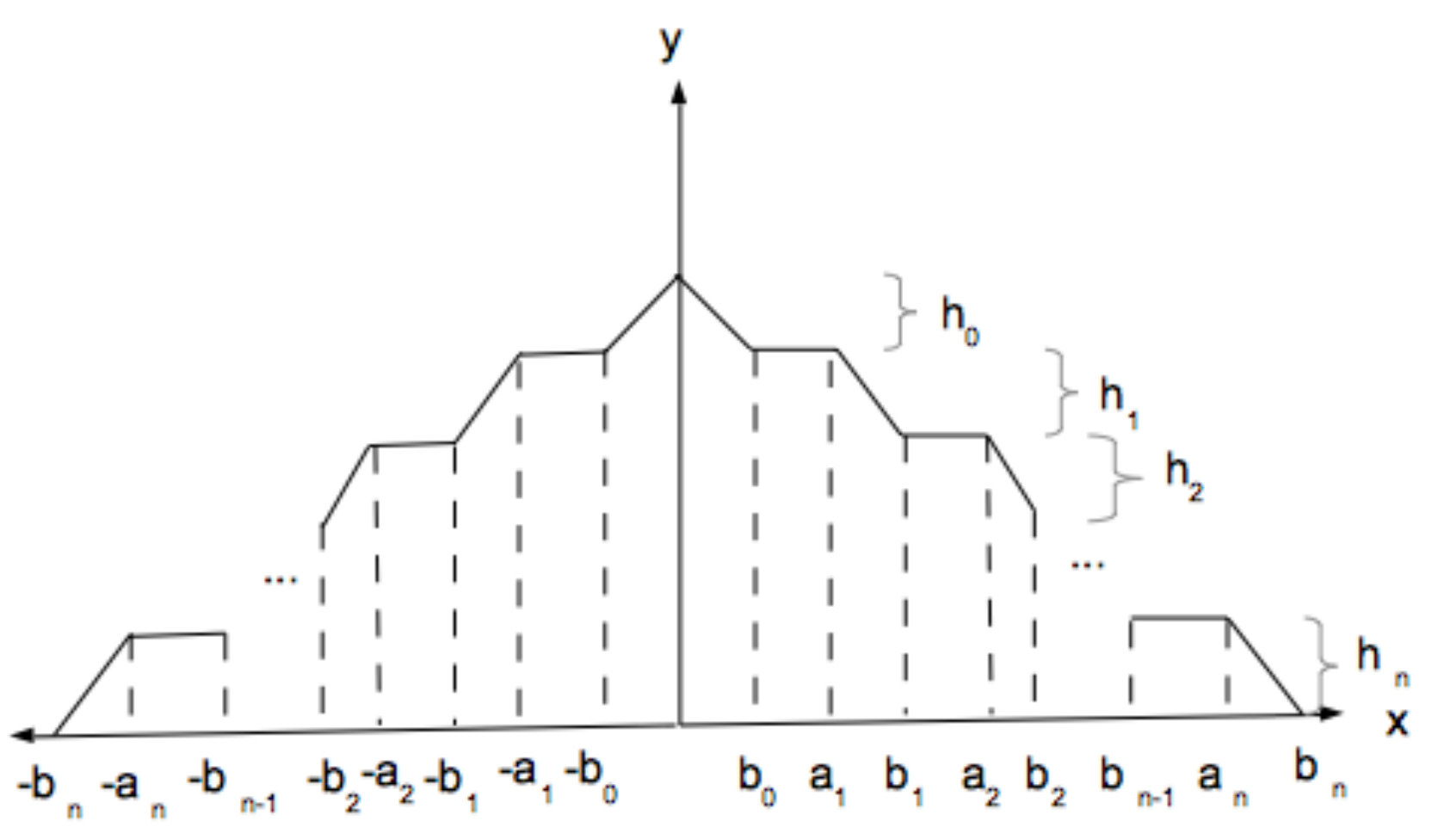}
	\caption{Emitter compounded of an arbitrary number, n, of trapezoidal protrusions of height $h_{j}$ and half-width from the lower base $b_{j}$ ($j \in \{1,2,...,n\}$) , with a triangular protrusion of height $h_{0}$ and half-width $b_{0}$ placed on the center of the top of the highest trapezoidal protrusion.}	\label{TrapTriang}
\end{figure}

Finally, the case of the emitter from the last section with a triangular protrusion placed on the center of the top of the highest trapezoidal stage will be considered, see Fig. \ref{TrapTriang}.  The Schwarz-Christoffel transformation mapping the u-axis in the w=u+iv$\rightarrow$(u,v)-plane into the polygonal line in Fig. \ref{TrapTriang} is given by the expression
\begin{equation} \label{SCT4}
z(w)=A \int_{u_0}^{w}  \frac{w^{1-\alpha}dw}{(w^{2}-1)^{\frac{1-\alpha}{2}}}  \prod_{j=1}^{n}\left[\frac{w^{2}-u_{j}^{2}}{w^{2}-v_{j}^{2}} \right]^{\alpha_{j}}+B,
\end{equation}
where $\alpha=\frac{2}{\pi} \arctan \left( \frac{b_0}{h_0} \right)$ corresponds to the internal angle from the upper vertex of the triangular protrusion, $\alpha_{j}=(1/ \pi) \arctan \left( \frac{h_j}{b_{j}-a_{j}} \right)$, $ \forall j \in \{1,2,...,n\}$ corresponds to the angles of the lower base of each trapeze with the adjacent sides and the parameters $u_j$ and $v_j$ satisfy the following correspondences ($j \in \{1,2,...,n \}$) between points in the w- and z-planes: $z(w=u_{j})=a_{j}+i \sum_{k=j}^{n} h_{k}$ and $z(w=v_{j})=b_{j}+i \sum_{k=j+1}^{n}h_{k}$. Based on the Riemann Mapping Theorem \cite{Riemann}, one can choose $z(w=u_{0} \equiv 0)=iH$ and $z(w=v_{0} \equiv 1)=b_{0}+i(H-h_{0})$, where $H \equiv \sum_{j=0}^{n} h_{j}$. The correspondences for $w=u_{0} \equiv 0$ and $w=v_{0} \equiv 1$ lead to the equations
\begin{align}
B=iH, \\
A =\frac{\sqrt{b_{0}^{2}+h_{0}^{2}}}{\int_{0}^{1}  \frac{w^{1-\alpha}dw}{(1-w^{2})^{\frac{1-\alpha}{2}}}  \prod_{j=1}^{n}\left[\frac{u_{j}^{2}-w^{2}}{v_{j}^{2}-w^{2}} \right]^{\alpha_{j}}}.
\end{align}
These results together with the remaining correspondences lead to the following equations $(j \geq 1)$:
\begin{equation} 
\frac{a_{j}-b_{j-1}}{\sqrt{b_{0}^{2}+h_{0}^{2}}}=\frac{\int_{v_{j-1}}^{u_{j}}  \frac{w^{1-\alpha}dw}{(w^{2}-1)^{\frac{1-\alpha}{2}}}  \prod_{k=1}^{j-1}\left[\frac{w^{2}-u_{k}^{2}}{w^{2}-v_{k}^{2}} \right]^{\alpha_{k}} \prod_{l=j}^{n}\left[\frac{u_{l}^{2}-w^{2}}{v_{l}^{2}-w^{2}} \right]^{\alpha_{l}}}{ \int_{0}^{1}  \frac{w^{1-\alpha}dw}{(1-w^{2})^{\frac{1-\alpha}{2}}}  \prod_{k=1}^{n}\left[\frac{u_{k}^{2}-w^{2}}{v_{k}^{2}-w^{2}} \right]^{\alpha_{k}}}  ,
\end{equation}
and
\begin{multline}
\frac{\sqrt{(b_{j}-a_{j})^{2}+h_{j}^{2}}}{\sqrt{b_{0}^{2}+h_{0}^{2}}}= \\
=\frac{\int_{u_{j}}^{v_{j}}  \frac{w^{1-\alpha}dw}{(w^{2}-1)^{\frac{1-\alpha}{2}}}  \left(\frac{w^{2}-u_{j}^{2}}{v_{j}^{2}-w^{2}} \right)^{\alpha_{j}}  \prod_{k=1}^{j-1}\left[\frac{w^{2}-u_{k}^{2}}{w^{2}-v_{k}^{2}} \right]^{\alpha_{k}}  \prod_{l=j+1}^{n}\left[\frac{u_{l}^{2}-w^{2}}{v_{l}^{2}-w^{2}} \right]^{\alpha_{l}} }{ \int_{0}^{1}  \frac{w^{1-\alpha}dw}{(1-w^{2})^{\frac{1-\alpha}{2}}}  \prod_{k=1}^{n}\left[\frac{u_{k}^{2}-w^{2}}{v_{k}^{2}-w^{2}} \right]^{\alpha_{k}} }.
\end{multline}
Combining the last equations, one obtains
\begin{multline} \label{Eqhv4}
\frac{\sqrt{(b_{j}-a_{j})^{2}+h_{j}^{2}}}{a_{j}-b_{j-1}}=\frac{h_{j}}{(a_{j}-b_{j-1}) \sin (\pi \alpha_{j})}= \\
=\frac{\int_{u_{j}}^{v_{j}}  \frac{w^{1-\alpha}dw}{(w^{2}-1)^{\frac{1-\alpha}{2}}}  \left(\frac{w^{2}-u_{j}^{2}}{v_{j}^{2}-w^{2}} \right)^{\alpha_{j}}  \prod_{k=1}^{j-1}\left[\frac{w^{2}-u_{k}^{2}}{w^{2}-v_{k}^{2}} \right]^{\alpha_{k}}  \prod_{l=j+1}^{n}\left[\frac{u_{l}^{2}-w^{2}}{v_{l}^{2}-w^{2}} \right]^{\alpha_{l}} }{\int_{v_{j-1}}^{u_{j}}  \frac{w^{1-\alpha}dw}{(w^{2}-1)^{\frac{1-\alpha}{2}}}  \prod_{k=1}^{j-1}\left[\frac{w^{2}-u_{k}^{2}}{w^{2}-v_{k}^{2}} \right]^{\alpha_{k}} \prod_{l=j}^{n}\left[\frac{u_{l}^{2}-w^{2}}{v_{l}^{2}-w^{2}} \right]^{\alpha_{l}}}.
\end{multline}
At this point, one may consider the limit $b_{0}<<a_{1} \leq b_{1}<<h_{1}<<a_{2}\leq b_{2}<<h_{2}<<...<<a_{n}\leq b_{n}<<h_{n} \Rightarrow 1<<u_{1}<<v_{1}<<u_{2}<<v_{2}<<...<<u_{n}<<v_{n}$ and then realize the following approximations, very similar to the ones from the previous section: 
\begin{align}
\frac{h_{j}}{(a_{j}-b_{j-1}) \sin (\pi \alpha_{j})} \approx \frac{h_{j}} {a_{j} \sin (\pi \alpha_{j})},
\end{align}
\begin{multline}
\int_{v_{j-1}}^{u_{j}}  \frac{w^{1-\alpha}dw}{(w^{2}-1)^{\frac{1-\alpha}{2}}}  \prod_{k=1}^{j-1}\left[\frac{w^{2}-u_{k}^{2}}{w^{2}-v_{k}^{2}} \right]^{\alpha_{k}} \prod_{l=j}^{n}\left[\frac{u_{l}^{2}-w^{2}}{v_{l}^{2}-w^{2}} \right]^{\alpha_{l}} \approx \\
\approx  \frac{1}{v_{j}^{2 \alpha_{j}}} \prod_{k=j+1}^{n} \left(\frac{u_k}{v_k} \right)^{\alpha_{k}} \int_{v_{j-1}}^{u_{j}} \frac{(u_{j}^{2}-w^{2})^{\alpha_{j}} w^{2\alpha_{j-1}}dw}{(w^{2}-v_{j-1}^{2})^{\alpha_{j-1}}} \approx  \\
\approx \frac{u_{j}^{2 \alpha_{j}+1}}{v_{j}^{2 \alpha_{j}}} \frac{ \Gamma(\alpha_{j}+1) \sqrt{\pi}}{2 \Gamma(\alpha_{j}+3/2)}\prod_{k=j+1}^{n} \left(\frac{u_k}{v_k} \right)^{\alpha_{k}},
\end{multline}
\begin{multline}
\int_{u_{j}}^{v_{j}}  \frac{w^{1-\alpha}dw}{(w^{2}-1)^{\frac{1-\alpha}{2}}}  \left(\frac{w^{2}-u_{j}^{2}}{v_{j}^{2}-w^{2}} \right)^{\alpha_{j}}  \prod_{k=1}^{j-1}\left[\frac{w^{2}-u_{k}^{2}}{w^{2}-v_{k}^{2}} \right]^{\alpha_{k}}  \prod_{l=j+1}^{n}\left[\frac{u_{l}^{2}-w^{2}}{v_{l}^{2}-w^{2}} \right]^{\alpha_{l}}  \\
\approx  \prod_{k=j+1}^{n} \left(\frac{u_k}{v_k} \right)^{\alpha_{k}} \int_{u_{j}}^{v_{j}} \left(\frac{w^{2}-u_{j}^{2}}{v_{j}^{2}-w^{2}} \right)^{\alpha_{j}} dw \approx \\
\approx \frac{v_{j} \Gamma(1-\alpha_{j}) \Gamma ( \alpha_{j}+1/2)}{\sqrt{\pi}}  \prod_{k=j+1}^{n} \left(\frac{u_k}{v_k} \right)^{\alpha_{k}} .
\end{multline}
After inserting the previous approximations and using the Gamma function reflection formula, Eq. (\ref{Eqhv4}) reduces to Eq. (\ref{Equv34}).

From Eq. (\ref{SCT4}), the FEF near the apex of the triangular protrusion $(w \approx 0)$ is given by;
\begin{equation} 
\gamma (w \approx 0) \approx  \frac{1}{|w|^{1-\alpha_{j}} } \prod_{j=1}^{n} \left(\frac{v_j}{u_j} \right)^{2 \alpha_{j}}.
\end{equation}
Besides that, Eq. (\ref{SCT4}) also leads to
\begin{equation} 
|z(w \approx 0)-iH| \approx \frac{A |w|^{2-\alpha}}{2-\alpha}  \prod_{j=1}^{n} \left(\frac{v_j}{u_j} \right)^{2 \alpha_{j}}.
\end{equation}
Thus, under the aforementioned approximations, the expression for the FEF in the vicinity of the apex of the emitter in Fig. \ref{TrapTriang} is given by
\begin{equation} \label{Final4}
\gamma (x \approx 0,y \approx H) \approx \left(\prod_{j=1}^{n} \gamma_{j} \right) \gamma_{T}, 
\end{equation}
where $\gamma_{j}$ and $\gamma_T$ are defined by the following equations:
\begin{align}
\gamma_{j} =\left[ \frac{ \alpha_{j} \left[\Gamma (\alpha_{j}) \right]^{2}}{(2 \alpha_{j}+1) \left[\Gamma \left(\alpha_{j}+\frac{1}{2} \right) \right]^{2} } \frac{h_j}{a_j}\right]^{2 \alpha_{j} /(2 \alpha_{j}+1)}, \\
\gamma_{T} = \left[ \frac{\sqrt{\pi}}{(2-\alpha) \frac{\xi(x,y)}{h_0}  \Gamma \left(1-\frac{\alpha}{2} \right) \Gamma \left(\frac{1+\alpha}{2} \right)}\right]^{\frac{1-\alpha}{2-\alpha}},
\end{align}
and $\xi(x,y)=\sqrt{x^{2}+(y-H)^{2}}$. Eq. (\ref{Final4}) shows that the FEF in the vicinity of the apex of the emitter in Fig. (\ref{TrapTriang}) is explicitly written as the product of the FEFs of each of the trapezoidal stages on a line $(\gamma_{j})$ and the FEF close to apex of a triangular protrusion on a line $(\gamma_{T})$. Thus, SC is proved. By considering $\alpha_{j}=1/2$ for all $j \in \{1,2,...,n\}$, the trapezes become rectangles and the result in Eq. (\ref{Final2}) is recovered. 

\section{Conclusions}

In the present work Schwarz-Christoffel conformal mapping \cite{SCT1,SCT2} was used to perform a completely analytical proof of SC for multi-stage field emitters in the limit in which each stage has much larger dimensions than the ones from the stages above. For simplicity of the calculations, only geometries with high aspect-ratio were considered, such that the FEFs corresponding to each of the single stages on a line can be evaluated by a simple analytical expression with no need of solving integral equations \cite{Miller1,Miller2}. The geometries considered here involve rectangular, trapezoidal and triangular shapes for the stages of the emitter and each stage is placed on the center of the top of another one. The number of stages is completely arbitrary.

The results obtained suggest the validity of SC for multi-stage field emitters with generic shape of the stages, even when there is lack of similarity between the stages, as long as the dimensions of any stage are much larger than the dimensions from the stages above and one stage is placed on the center of the top of another one. Moreover, this is the first analytical proof of SC for an arbitrary number of stages. 

As a limitation of the Schwarz-Christoffel technique, only ridge emitters are considered along this work. Nevertheless, the theoretical conclusions obtained from the completely analytical results derived along this manuscript are expected to be valid in a much broader sense.

\section*{Acknowledgement}
This study was financed in part by the Coordena\c{c}\~{a}o de Aperfei\c{c}oamento de Pessoal de N\'{i}vel Superior (CAPES), Finance Code 001. The author thanks Thiago Albuquerque de Assis for interesting discussions about the theme during the preparation of the present work.


\begin{thebibliography}{99}
	
	\bibitem{Jeffreys} H. Jeffreys, Proc. London Math. Soc. s2-23, 428 (1925).
	
	\bibitem{FowlerNordheim} R. H. Fowler and L. Nordheim, Proc. R. Soc. London, Ser. A 119, 173 (1928).
	
	\bibitem{Burgess} R. E. Burgess, H. Kroemer, and J. M. Houston, Phys. Rev. 90, 515 (1953).
	
	\bibitem{MurphyGood} E. L. Murphy and R. H. Good, Phys. Rev. 102, 1464 (1956).

	\bibitem{ForbesFundmentals} A. Fischer, M. S. Mousa and A. R. G. Forbes, J. Vac. Sci. Technol., B 31, 032201 (2013).
	
	\bibitem{JensenPRST} K. L. Jensen, D. A. Shiffler, J J. Petillo, Z. Pan, and J. W. Luginsland, Phys. Rev. ST Accel. Beams 17, 043402  (2014).

	\bibitem{Holgate} J. T. Holgate and M. Coppins, Phys. Rev. Applied 7, 044019 (2017).
		
	\bibitem{Nanomaterials} F. Urban, M. Passacantando, F. Giubileo, L. Lemmo and A. Di Bartolomeo, Nanomaterials 2018, 8(3), 151.
	
	\bibitem{Mueller1} E. W. M\"{u}ller, Z. Phys. 106, 541 (1937).
	
	\bibitem{Mueller2} E. W. M\"{u}ller, Z. Phys. 131, 136 (1951).
	
	\bibitem{Mueller3} E. W. M\"{u}ller and K. Bahadur, Phys. Rev. 102, 624 (1956).
			
	\bibitem{Ultramicroscopy} R. G. Forbes, C. Edgcombe, and U. Valdre, Ultramicroscopy \textbf{95}, 57-65 (2003).

	\bibitem{ColeBook} M. T. Cole, M. Mann, K. B. Teo, and W. I. Milne, in Emerging Nanotechnologies for Manufacturing, Micro and Nano Technologies, 2nd ed., edited by W. Ahmed and M. J. Jackson (William Andrew Publishing, Boston, 2015).
	
	\bibitem{ForbesCFE1} R. G. Forbes and J. H. Deane, Proc. R. Soc. London, Ser. A 463, 2907 (2007).
	
	\bibitem{ForbesCFE2} R. G. Forbes, Proc. R. Soc. London, Ser. A 469, 20130271 (2013).
	
	\bibitem{Colgan} M. Colgan and M. Brett, Thin Solid Films 389 (2001).
			
	\bibitem{Xu} N. Xu and S. E. Huq, Mater. Sci. Eng. R 48 47-189 (2005).

	\bibitem{ForbesCFE3} R. G. Forbes, Nanotechnology 23 095706 (2012).

	\bibitem{Han} J. W. Han, D. I. Moon and M. Meyyappan, Nano Lett.17 2146-51 (2017).
	
	\bibitem{MarcelinoJVSTB} E. Marcelino, T. A. de Assis, and C. M. C. de Castilho, J. Vac. Sci. Technol., B \textbf{35}, 051801 (2017).	

	\bibitem{MarcelinoJAP} E. Marcelino, T. A. de Assis, and C. M. C. de Castilho, J. Appl. Phys. \textbf{123}, 124302 (2018).
	
	\bibitem{MarcelinoPRApplied} E. Marcelino, T. A. de Assis, C. M. C. de Castilho and R. F. S. Andrade, Phys. Rev. Applied 11 014012 (2019).
	
	\bibitem{STFE1} T. A. de Assis and F. Dall' Agnol, Nanotechnology 27 44LT01 (2016).
	
	\bibitem{STFE2} T. A. de Assis and F. Dall' Agnol, J. Appl. Phys. 121 014503 (2017).
	
	\bibitem{STFE3} P. D. Joshi, D. S. Joaq, D. J. Late and I. S. Mulla, J. Vac. Sci. Technol. B 35 02C105 (2017).
	
	\bibitem{STFE4} A. Venkattraman, J. Phys. D: Appl. Phys. \textbf{47}, 425205 (2014).
	
	\bibitem{Schottky} W. Schottky, Z. Phys. 14, 63 (1923).
	
	\bibitem{Miller1} R. Miller, Y. Y. Lau, and J. H. Booske, Appl. Phys. Lett. \textbf{91}, 074105 (2007).
	
	\bibitem{Miller2} R. Miller, Y. Y. Lau, and J. H. Booske, J. Appl. Phys. \textbf{106}, 104903 (2009).
	
	\bibitem{WangAPL2004} S. H. Jo, D. Z. Wang, J. Y. Huang, W. Z. Li, K. Kempa, and Z. F. Ren, Appl. Phys. Lett. \textbf{85}, 810  (2004).
	
	\bibitem{HuangTEM} J. Y. Huang, K. Kempa, S. H. Jo, S. Chen, and Z. F. Ren, Appl. Phys. Lett. \textbf{87}, 053110  (2005). 

	\bibitem{AIPJensen}  K. L. Jensen, D. A. Shiffler, J. R. Harris, and J. J. Petillo, AIP Advances \textbf{6}, 065005 (2016).
		
	\bibitem{Jensen_Verif} J. R. Harris and J. W. Lewellen, J. Appl. Phys. 125, 215306 (2019).
	
	\bibitem{Jensen_Invest} J. R. Harris, D. A. Shiffler, K. L. Jensen , and J. W. Lewellen, J. Appl. Phys. 125, 215307 (2019).
		
	\bibitem{SCT1} J. Brown and R. Churchill, Complex Variables and Applications, Brown and Churchill Series (McGraw-Hill Higher Education, New York, 2013).
	
	\bibitem{SCT2} F. Hildebrand, Advanced Calculus for Applications (Prentice-Hall, NJ, 1962).
	
	\bibitem{Riemann} B. Riemann, “Grundlagen für eine allgemeine Theorie der Functionen einer veränderlichen complexen Grösse”, Ph.D. thesis, University of Göttingen, 1851.

	\bibitem{MarcelinoErratum} E. Marcelino, T. A. de Assis, and C. M. C. de Castilho, J. Appl. Phys. 124, 159901 (2018).






	
	
	
	
	
	
		
		
	
	
	
	
	
		
	
	
	
	
	
	
	
	
	
	
		
	
	
	
	
	
	
	
	
	
	
	
	
	
	
	
	
	

	
		

	
	

	

	
	
	
	




	
	

	


	

	
	

		
\end{thebibliography}

\end{document}